\documentstyle[11pt,paspconf,164page,164def,psfig]{article}
\begin{document}

\title{GRS\,1915+105: Flares, QPOs and other events at 15\,GHz}

\index{GRS\,1915+105|(}
\index{X-ray transient|(}
\index{quasi-periodic oscillations|(}

\label{poole}

\markboth{Pooley \& Fender}{GRS\,1915+105}

\author{G. G. Pooley}
\affil{Mullard Radio Astronomy Observatory, Cavendish Laboratory, Cambridge, UK}

\author{R. P. Fender}
\affil{Astronomy Centre, University of Sussex, UK}

\begin{abstract}
Monitoring with the Ryle Telescope at 15\,GHz of the Galactic X-ray transient
source GRS\,1915+105 has revealed a remarkable range of rapid and
extended flares which appear to be related to the X-ray emission
as recorded by the {\it RXTE} all-sky monitor. Quasi-periodic oscillations
in the range 20 -- 40 min have been found and are probably related to oscillations
in the soft X-ray flux.
\end{abstract}

\section{Introduction}

The Galactic X-ray transient GRS\,1915+105 has proved to have a rich
structure in its high- and low-energy X-ray emission. In the radio regime,
it is no less remarkable: Mirabel \& Rodr\'\i guez (1995) discovered
a double-sided relativistic ejection for which they derive a velocity
of 0.92$c$. The distance is estimated, from H\,{\sc i} absorption measurements,
to be 12.5 kpc. Radio monitoring had already shown the emission to be
highly variable (Rodr\'\i guez et al.\ (1995), Foster et al.\ (1996)),
and we started monitoring the source at 15 GHz in mid-1995.
The observational details and further results are described by
Pooley \& Fender (1997).

\section{Results}

Figure 1 shows the data for some 17 months. Individual observations lasted
from less than 1 hour to about 6 hours. It can be seen that the variations
are unpredictable and frequently very rapid; the flux density can increase
from less than 1 mJy to 100 mJy and back in less than a day. The major flare
event starting in 1996 July was characterised by relatively smooth variations
during the first month or so; subsequently the source became highly variable
and showed frequent examples of quasi-periodic variations.

\begin{figure}[h]
\cl{\psfig{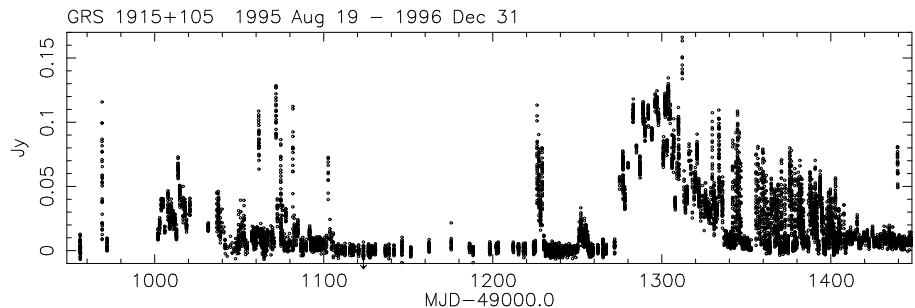}}
\caption{Flux density of GRS\,1915+105 at 15\,GHz over 17 months.
Each point plotted is a 5-min integration.\label{poolef1}}
\end{figure}

\subsection{Quasi-periodic oscillations}

A characteristic form of the outbursts observed at 15 GHz is an event with a
rise-time of less than 5 min, followed by a roughly exponential decay with
time constant between 12 and 25 min. These may be isolated, they may recur
with apparently random intervals, or they may repeat with some regularity;
the QPOs seem to favour intervals near 25 and 40 minutes, although not
exclusively so. Some examples are shown in Figure 2.

\begin{figure}[h]
\cl{\psfig{file=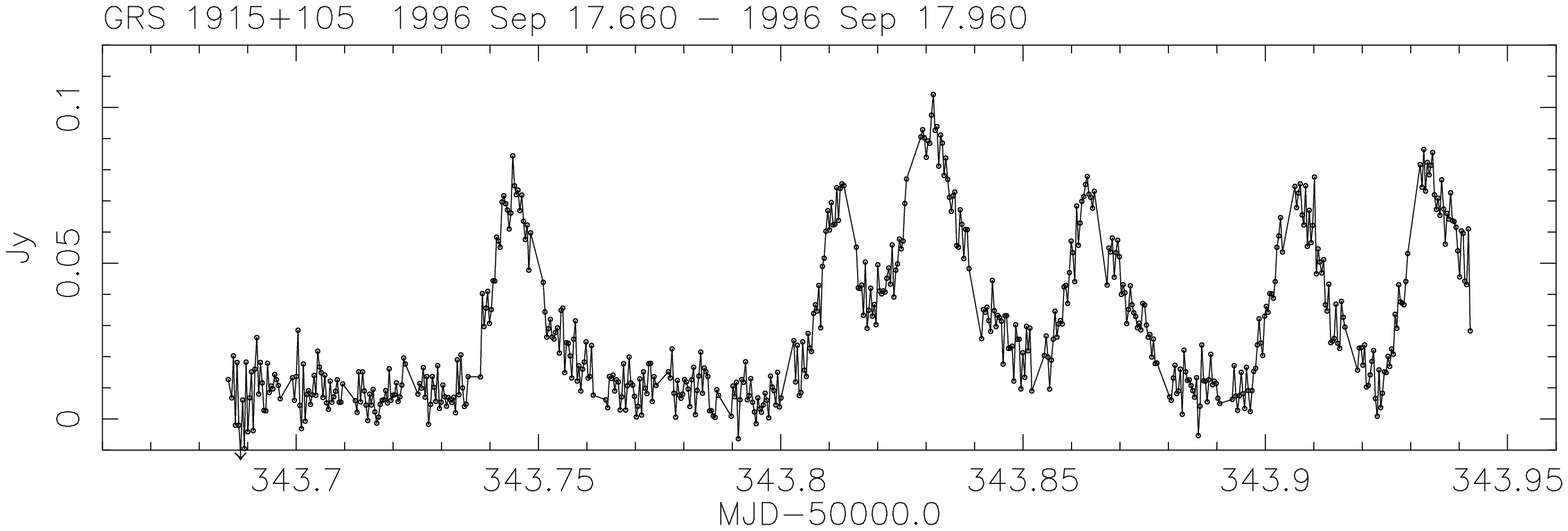,angle=270,width=0.5\textwidth}\quad\psfig{file=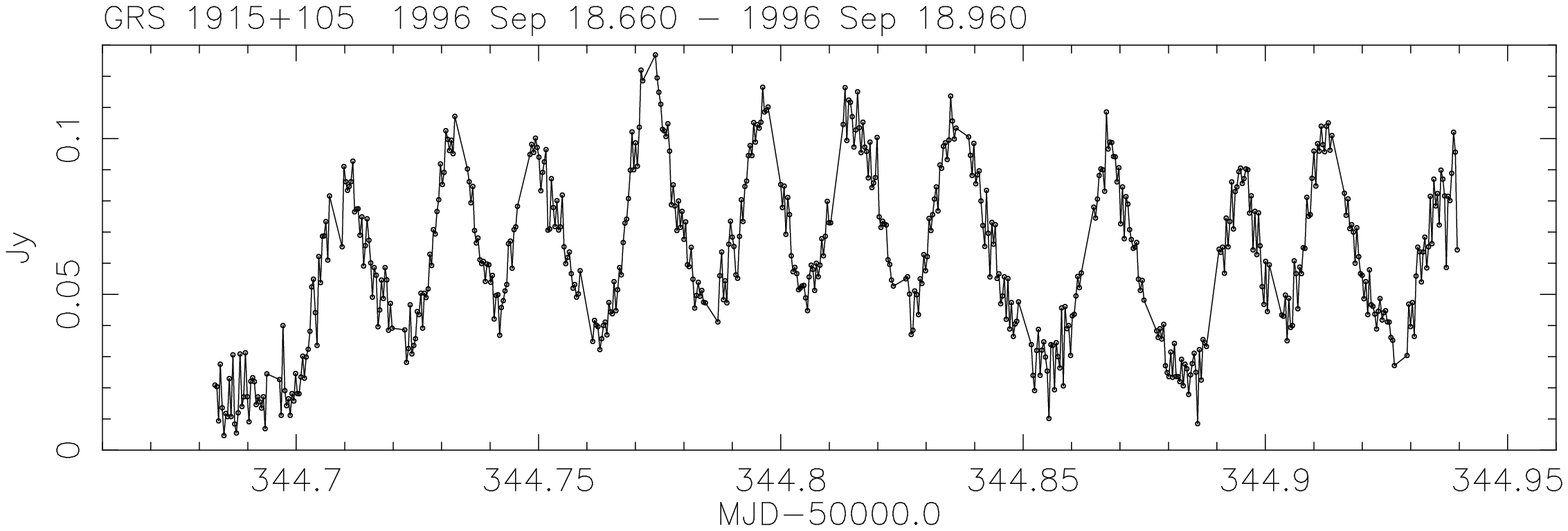,angle=270,width=0.5\textwidth}}
\vspace{2mm}
\cl{\psfig{file=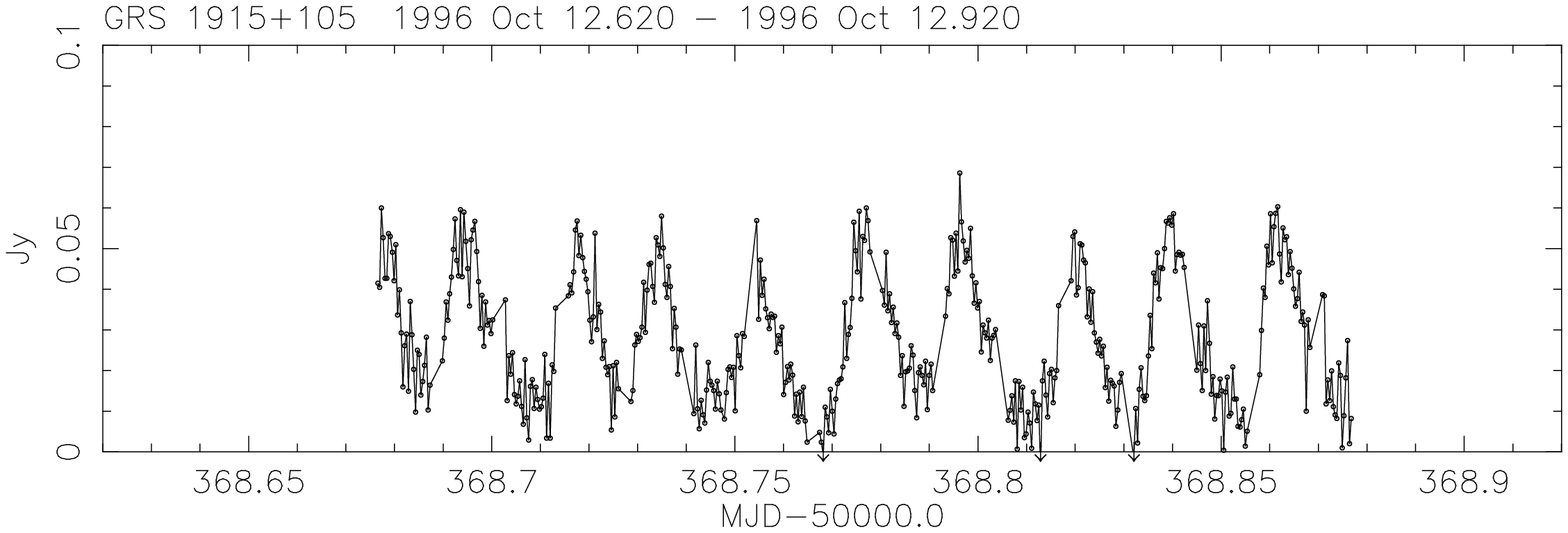,angle=270,width=0.5\textwidth}\quad\psfig{file=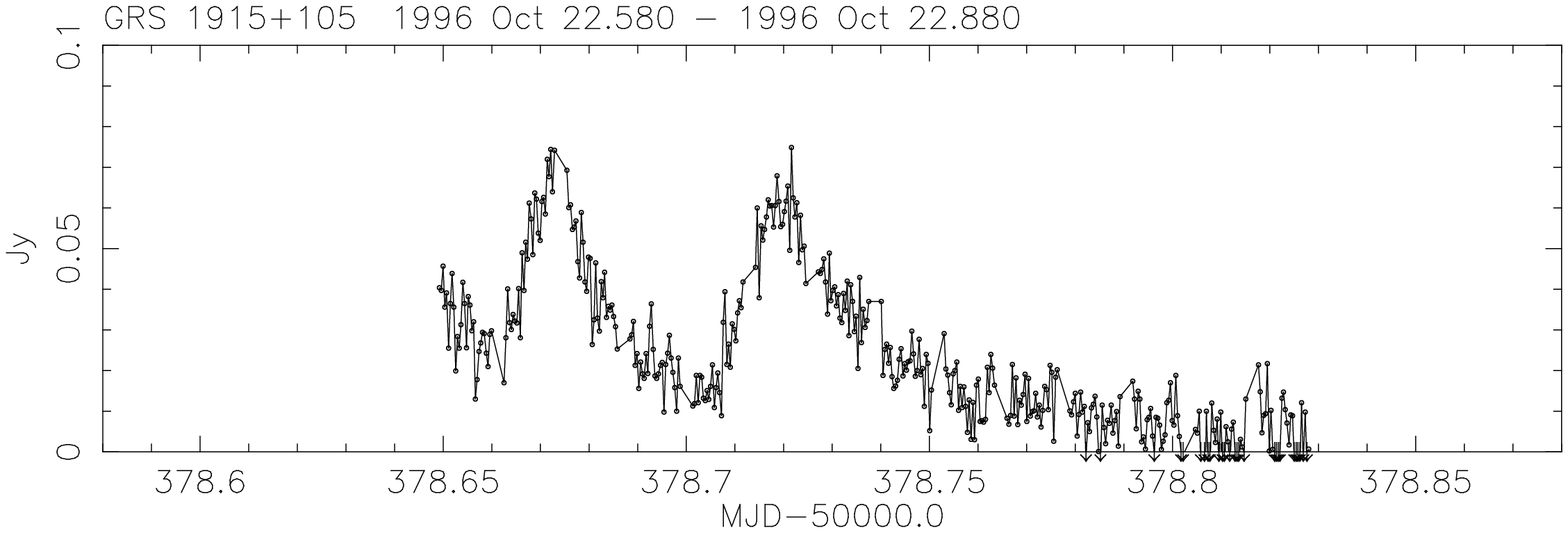,angle=270,width=0.5\textwidth}}
\caption{QPOs in GRS\,1915+105 at 15\,GHz; the integration time is 32\,s.\label{poolef2}}
\end{figure}

\subsection{Comparison with other wavebands}

The pattern of radio emission is related in a rather complex way to
that of the soft X-ray, as recorded by the {\it RXTE} all-sky monitor.
Active (rapidly-varying) X-ray emission is usually, but not always,
accompanied by radio emission. During the 1996 July radio flare,
the X-ray emission was unusually constant.

Obtaining simultaneous observations of the rapid QPO events is
difficult, since they are very unpredictable, but we believe
that these observations will be necessary for a better understanding
of the source. Perhaps the most
intriguing simultaneous observation is that on 1996 Oct 24, when
the PCA on the {\it RXTE} satellite detected oscillations which appear,
on the basis of only a few cycles, to be related in phase with the 15-GHz data.

One observation with the VLBA at 8 GHz, on 1996 May 24, overlapped with
a Ryle Telescope observation; the variations at 15 GHz appear about
4 minutes earlier than those at 8 GHz (Dhawan, \tp, \page \pageref{dhawa}).

Fender et al. (1997) have observed outbursts in the infrared with
time-scales and flux densities similar to those of the
radio events. We do not yet have simultaneous IR and radio data for
these flares.

\index{GRS\,1915+105|)}
\index{X-ray transient|)}
\index{quasi-periodic oscillations|)}

\end{document}